\documentclass[fleqn,10pt]{wlscirep}
\usepackage[utf8]{inputenc}
\usepackage[T1]{fontenc}
\usepackage{changes}
\usepackage{figcaps}
\usepackage[english]{babel}

\usepackage{amsmath,fixmath}
\usepackage{lineno}
\usepackage[explicit]{titlesec}

\usepackage{graphicx}
\usepackage[labelfont=bf]{caption}
\usepackage[format=hang]{subcaption}

\usepackage{tabularx}
\usepackage[justification=justified, format=plain]{caption}

\usepackage{algorithm,algorithmic}

\hypersetup{
 bookmarks=true, 
 unicode=false, 
 pdftoolbar=true, 
 pdfmenubar=true, 
 pdffitwindow=false, 
 pdfstartview={FitH}, 
 pdftitle={}, 
 pdfauthor={}, 
 pdfcreator={}, 
 pdfproducer={}, 
 pdfkeywords={} {} {}, 
 pdfnewwindow=true, 
 colorlinks=true, 
 linkcolor=red, 
 filecolor=magenta, 
 urlcolor=cyan 
}

\newcommand{\be}{\begin{equation}}
\newcommand{\ee}{\end{equation}} 
\newcommand{\bea}{\begin{eqnarray}}
\newcommand{\eea}{\end{eqnarray}}

\newcommand{\grad}{{\nabla}} 

\renewcommand{\v}[1]{\mathbf{#1}} 

\newcommand{\f}[2]{\frac{#1}{#2}}

\newcommand{\ccup}[1]{\left\{#1\right\}}

\newcommand{\bup}[1]{\left(#1\right)}
\newcommand{\rup}[1]{\left[#1\right]}

\renewcommand{\t}[1]{\tilde{#1}}

\newcommand{\s}{\sigma}

\renewcommand{\ref}[1]{[\ref{#1}]}

\graphicspath{{./figures/}} 

\usepackage[normalem]{ulem} 

\newcommand{\nix}[1]{\sout{#1}}

\newcommand{\dmkS}{\mbox{{\small \textit{DMK-Solver}}}}
\newcommand{\dmk}{\mbox{{\small DMK}}}
\newcommand{\ddmk}{\mbox{{\small \textit{discrete-DMK-Solver}}}}

\begin{document}

\title{Network extraction by routing optimization}

\author[1,+]{Diego Baptista}
\author[1,+]{Daniela Leite}
\author[2]{Enrico Facca}
\author[3]{Mario Putti}
\author[1,*]{Caterina De Bacco}
\affil[1]{Max Planck Institute for Intelligent Systems, Cyber Valley, 72076, Tübingen, Germany}
\affil[2]{Centro di Ricerca Matematica Ennio De Giorgi, Scuola Normale Superiore, Piazza dei Cavalieri, 3, Pisa, Italy}
\affil[3]{Department of Mathematics “Tullio Levi-Civita”, University of Padua, via Trieste 63, Padua, Italy}

\affil[*]{caterina.debacco@tuebingen.mpg.de}
\affil[+]{These authors contributed equally to this work}
\figcapsoff
\printfigures

\begin{abstract}
Routing optimization is a relevant problem in many contexts. Solving directly this type of optimization problem is often computationally unfeasible. Recent studies suggest that one can instead turn this problem into one of solving a dynamical system of equations, which can instead be solved efficiently using numerical methods. This results in enabling the acquisition of optimal network topologies from a variety of routing problems. However, the actual extraction of the solution in terms of a final network topology relies on numerical details which can prevent an accurate investigation of their topological properties. In fact, in this context, theoretical results are fully accessible only to an expert audience and ready-to-use implementations for non-experts are rarely available or insufficiently documented.  
In particular, in this framework, final graph acquisition is a challenging problem in-and-of-itself. Here we introduce a method to extract networks topologies from dynamical equations related to routing optimization under various parameters' settings. Our method is made of three steps: first, it extracts an optimal trajectory by solving a dynamical system, then it pre-extracts a network and finally, it filters out potential redundancies. Remarkably, we propose a principled model to address the filtering in the last step, and give a quantitative interpretation in terms of a transport-related cost function. This principled filtering can be applied to more general problems such as network extraction from images,  thus going beyond the scenarios envisioned in the first step.
Overall, this novel algorithm allows practitioners to easily extract optimal network topologies by combining basic tools from numerical methods, optimization and network theory. Thus, we provide an alternative to manual graph extraction which allows a grounded extraction from a large variety of optimal topologies. The analysis of these may open up the possibility to gain new insights into the structure and function of optimal networks. Our algorithm is open source and available at \url{https://github.com/Danielaleite/NextRout}.
\end{abstract}

\flushbottom
\maketitle

\thispagestyle{empty}


\section*{Introduction}
\label{sec:introduction}

Investigating optimal network topologies is a relevant problem in several contexts, with applications varying from transportation networks\cite{banavar2000topology,corson2010fluctuations,li2010towards,yeung2013physics}, communication systems\cite{guimera2002optimal,donetti2005entangled,ronellenfitsch2018optimal}, biology \cite{gazit1995scale,garlaschelli2003universal} and ecology\cite{balister2018river,santambrogio2007optimal,katifori2010damage}. Depending on the specified objective function and set of constraints of a routing optimization problem \cite{santambrogio2015optimal}, optimal network topologies can be determined by different processes ranging from energy-minimizing tree-like structures  ensuring steeper descent through a landscape as in river basins \cite{balister2018river} to the opposite scenario of loopy structures that favor robustness to fluctuations and damage as in leaf venation \cite{messinger2007task,katifori2010damage}, the retina vascular system \cite{fruttiger2002development,schaffer2006two} or noise-cancelling networks \cite{ronellenfitsch2018optimal}.\\
In many applications, optimal networks can arise from an underlying process defined on a continuous space rather than a discrete network as in standard combinatorial optimization routing problems \cite{altarelli2015edge,bayati2008statistical,de2014shortest,braunstein2018cavity}.
Optimal routing networks try to move resources by minimizing the transportation cost. This cost may be taken to be a function of the traveled distance, such as in Steiner trees, or proportional to the dissipated energy, such as optimal channel networks or resistance networks. The common denominator of these configurations is that they have a tree-like shape, i.e., optimal routing networks are loopless\cite{banavar2000topology,Ronellenfitsch2016}. Recent developments in the mathematical theory of optimal transport~\cite{santambrogio2007optimal,santambrogio2015optimal} have proved that this is indeed the case and that complex fractal-like networks arise from branched optimal transport problems~\cite{Brancolini2014}. While the theory starts to consolidate, efficient numerical methods are still in a pre-development stage, in particular in the case of branched transport, where only a few results are present~\cite{Xia2015,Pegon2019}, reflecting the obstacle that all these problems have an NP-hard genesis.
Recent promising results \cite{facca2016towards,facca2019numerics} map a computationally hard optimization problem into finding the long-time behavior of a system of dynamic partial differential equations, the so-called Dynamic Monge-Kantorovich (DMK) approach, which is instead numerically accessible, computationally efficient, and leads to network shapes that resemble optimal structures~\cite{facca2020branch}. Working in discretized continuous space, and in many network-based discretizations such as lattice-like networks as well, requires the use of threshold values for the identification of active network edges. This has the main consequence that there might be no obvious final resulting network, an output that would be trivial when starting from an underlying search space formed by predefined selected network structures. For example, the output of a numerically discretized (by, e.g., the Finite Element method) routing optimization problem in a 2D space is a real-valued function on a set of $(x,y)$ points defined on a grid or triangulation, which already has a graph structure. Despite the underlying graph, this grid function contains numerous side features, such as small loops and dangling vertices, that prevent the recognition of a clear optimal network structure. Obtaining this requires a suitable identification of vertices and edges that should contain the optimal network properties embedded in the underlying continuous space. In other words, the output of a routing optimization problem in continuous space carries unstructured information about optimality that is hard to interpret in terms of network properties. Extracting a network topology from this unstructured information would allow, on one hand, better interpretability of the solution and enable the comparison with networks resulting from discrete space. On the other hand, the use of tools from network theory to investigate the properties related to optimality, for instance, to perform clustering or classification tasks based on a set of network features.
  One can frame this problem as that of properly compressing the information contained in the ``raw'' solution of a routing problem in continuous space into an interpretable network structure while preserving the important properties connected to optimality. This is a challenging task, as compression might result in losing important information. The problem is made even more complex because one may not know in advance what are the relevant properties for the problem at hand, a knowledge that could help drive the network extraction procedure. This is the case for any real network, where the intrinsic optimality principle is elusive and can only be speculated about by observing trajectories, an approach adopted for instance when processing images in biological networks\cite{baumgarten2010detection,obara2012bioimage,bebber2007biological,boddy2010fungal}.

Several works have been proposed to tackle domain-specific network extraction.
These methods include  using segmentation techniques on a set of image pixels to extract a skeleton  \cite{obara2012bioimage, baumgarten2010detection,dehkordi2011review} that is then converted into a network; a pipeline combining different segmentation algorithms building from OpenCV, \cite{bradski2008learning} which is made available with an intuitive graphical interface \cite{dirnberger2015nefi}; graph-based techniques \cite{chai2013recovering} that sample junction-points from input images; methods that use deep convolutional neural networks \cite{wang2015road} or   minimum cost path computations \cite{wegner2015road} to extract road networks from images.
 These are mainly using image processing techniques as the input is an image or photograph, which might not necessarily be related to a routing optimization problem. 

In this work, we propose a new approach for the extraction of network topologies and build a protocol to address this problem. This can take in input the numerical solution of a routing optimization problem in continuous space as described in \cite{facca2016towards,facca2019numerics,facca2020branch} and then processes it to finally output the corresponding network topology in terms of a weighted adjacency matrix. However, it can also be applied to more general inputs, such as images, which may not necessarily come from the solution of an explicit routing transportation problem. 
Specifically, our work features a collection of numerical routines and graph algorithms designed to extract network structures that can then be properly analyzed in terms of their topological properties. The extraction pipeline consists in a sequence of three main algorithmic steps: i) compute the steady state solutions of the DMK equations (\dmkS); ii) extract the optimal network solution of the routing optimization problem (\textit{graph pre-extraction}); iii) filter the network removing redundant structure (\textit{graph filtering}).
While for this work we test and demonstrate our algorithm on routing scenarios coming from DMK, which constitute our main motivation, we remark that only the first step is specific to these, whereas the last two steps are applicable beyond these settings. The graph pre-extraction step consists of a set of rules aiming at building a network from an input that is not explicitly a topological structure made of nodes and edges. 
The filtering step is based on a principled mathematical model inspired by that of the first step, which leads to an efficient algorithmic implementation. Our network filter has a nice interpretation in terms of a cost function that interpolates between an operating cost and an infrastructure one, contrarily to common approaches used in image processing for filtering, which often relies on heuristics.

 A successful execution will return a representation of the network in terms of an edge-weighted undirected network. The resulting weights are related to the optimal flow, solution of the routing problem. Once the network is obtained, practitioners can deploy arbitrary available network analysis software\cite{hagberg2008exploring,xu2010comparative,batagelj1998pajek,bastian2009gephi} or custom-written scripts to investigate properties of the optimal topologies.
While our primary goal is to provide a framework and tool to solve the research question of how to extract network topologies resulting from routing optimization problems in continuous space or any other image containing a network structure, we also aim at encouraging non-expert practitioners to automatically extract networks from such problems or from more general settings beyond that. Thus we make available an open-source algorithmic implementation and executables of this work at \url{https://github.com/Danielaleite/Nextrout}.

  

\section{The routing optimization problem}\label{sec:routingopt}
In this section, we describe the main ideas and establish notation. We start by introducing the dynamical system of equations corresponding to the DMK routing optimization problem as proposed by Facca et al. \cite{facca2016towards,facca2019numerics,facca2020branch} In these works, the authors first generalize the discrete dynamics of the slime mold \textit{Physarum Polycephalum} (PP) to a continuous domain; then they conjecture that, like its discrete counterpart, its solution tends to an equilibrium point which is the solution of the Monge-Kantorovich optimal mass transport \cite{evans1999differential} as time goes to infinity.

We denote the space where the routing optimization problem is set as $\Omega \in \mathbb{R}^{n}$, an open bounded domain that compactly contains $f(x)=f^+(x)-f^-(x)\in \mathbb{R}$, the forcing function, describing the flow generating sources $f^+(x)$ and sinks $f^-(x)$. It is assumed that the system is isolated, i.e., no fluxes are entering or exiting the domain from the boundary. This imposes the constraint $\int_\Omega f(x)dx  = 0$ to ensure mass balance. It is assumed that the flow is governed by a transient Fick-Poiseuille type flux $q=- \mu \grad u$, where $\mu(t,x),u(t,x)$ are called \textit{conductivity} or \textit{transport density} and \textit{transport potential}, respectively. 

The continuous set  dynamical Monge-Kantorovich (DMK) equations are given by:
\bea\label{eqn:ddmk}
-\nabla \cdot (\mu(t,x)\nabla u(t,x)) &=& f^+(x)-f^-(x) \,, \label{eqn:ddmk1}\\
\frac{\partial \mu(t,x)}{\partial t}  &=& \rup{\mu(t,x)\nabla u(t,x)}^{\beta} - \mu(t,x) \,, \label{eqn:ddmk2}\\
\mu(0,x)  &=& \mu_0(x) > 0  \label{eqn:mu0} \,,
\eea
where $\nabla=\nabla_{x}$. Equation~(\ref{eqn:ddmk1}) states the spatial balance of the Fick-Poiseuille flux and is complemented by no-flow Neumann boundary conditions; Eq. (\ref{eqn:ddmk2}) enforces the system dynamics in analogy with the discrete PP model and Eq. (\ref{eqn:mu0}) provides the initial configuration of the system. The parameter $\beta$  captures different routing transportation mechanisms. A value of $\beta<1$ enforces optimal solutions to avoid traffic congestion; $\beta = 1$ is shortest path-like; while $\beta > 1 $ encourages consolidating the flow so to use a smaller amount of network-like infrastructure, and is related to branched transport \cite{xia2014landscape,santambrogio2007optimal}.
Within a network-like interpretation, qualitatively, $\mu(x,t)$ describes the capacity of the network edges\cite{hydraulic}. Thus, its initial distribution $\mu_{0}(x)$ describes how the initial capacities are distributed. 

\begin{figure}[htb]
\caption{Different values of $\beta$ in Eq. (\ref{eqn:ddmk2}) lead to different settings of a routing optimization problem. Colors denote different intensities of conductivity $\mu$ as described by the color bar on the left. $(a)$ $\beta$ < 1 enforces avoiding mass congestion; $(b)$ $\beta = 1$ is shortest path-like,  the mass goes  straight  from source to sink; $(c)$ $\beta > 1$ encourages traffic consolidation. Red rectangle denotes the sink, green ones the four sources.}
 \label{fig:beta}
\begin{subfigure}{0.33\textwidth}
\includegraphics[width=0.9\textwidth]{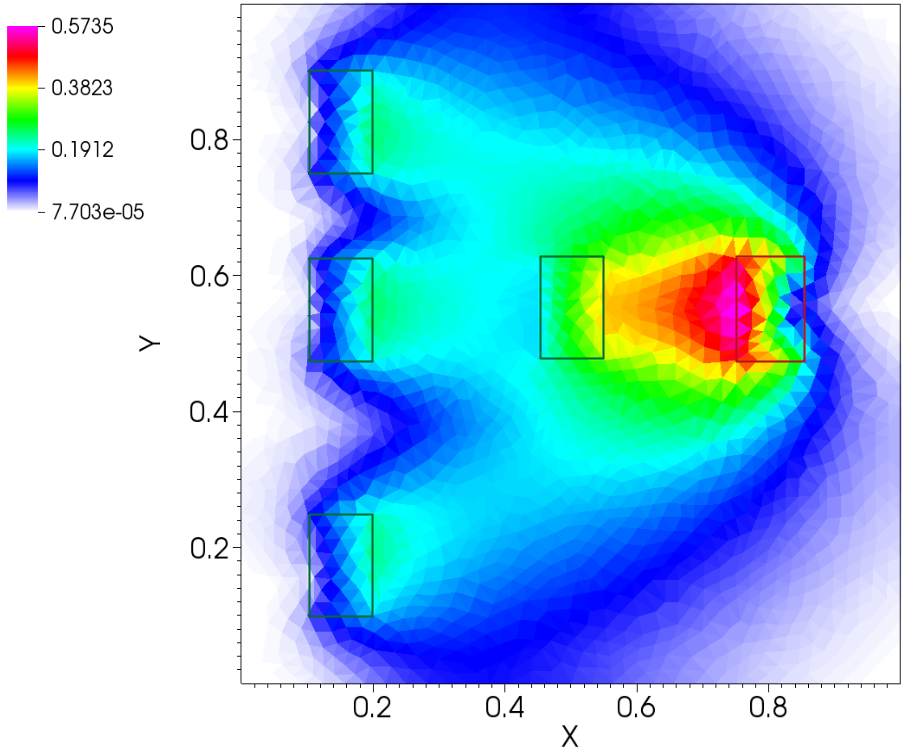} 
\caption{$\beta = 0.75$}
\label{fig:subim1}
\end{subfigure}
\begin{subfigure}{0.33\textwidth}
\includegraphics[width=0.9\textwidth]{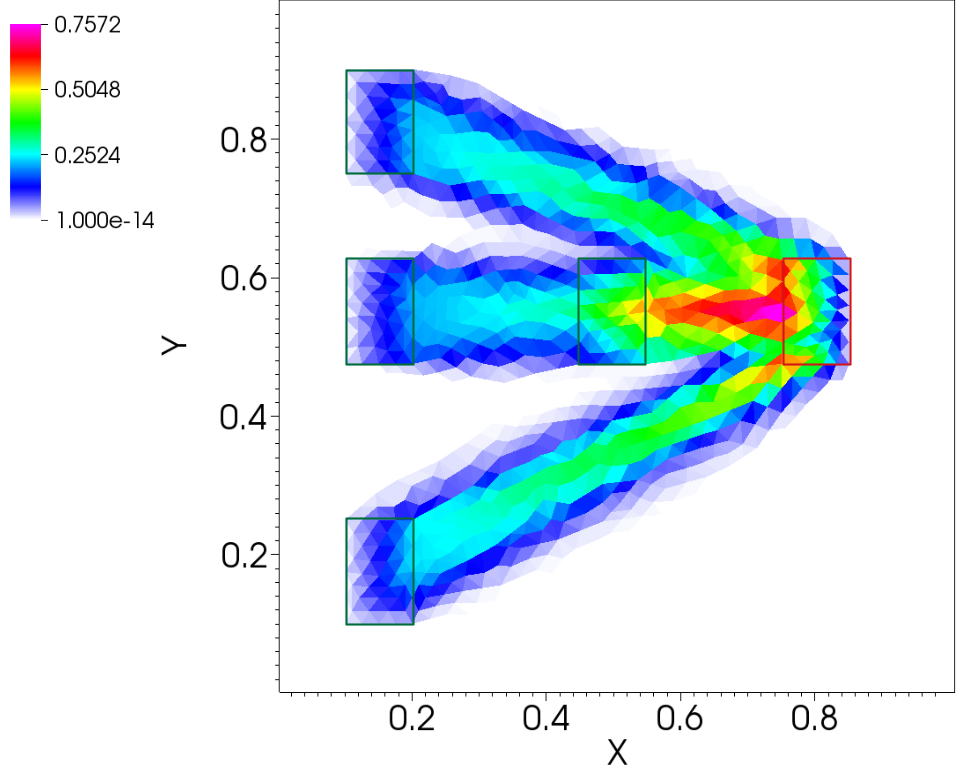}
\caption{$\beta = 1.0$}
\label{fig:subim2}
\end{subfigure}
\begin{subfigure}{0.33\textwidth}
\includegraphics[width=0.9\textwidth]{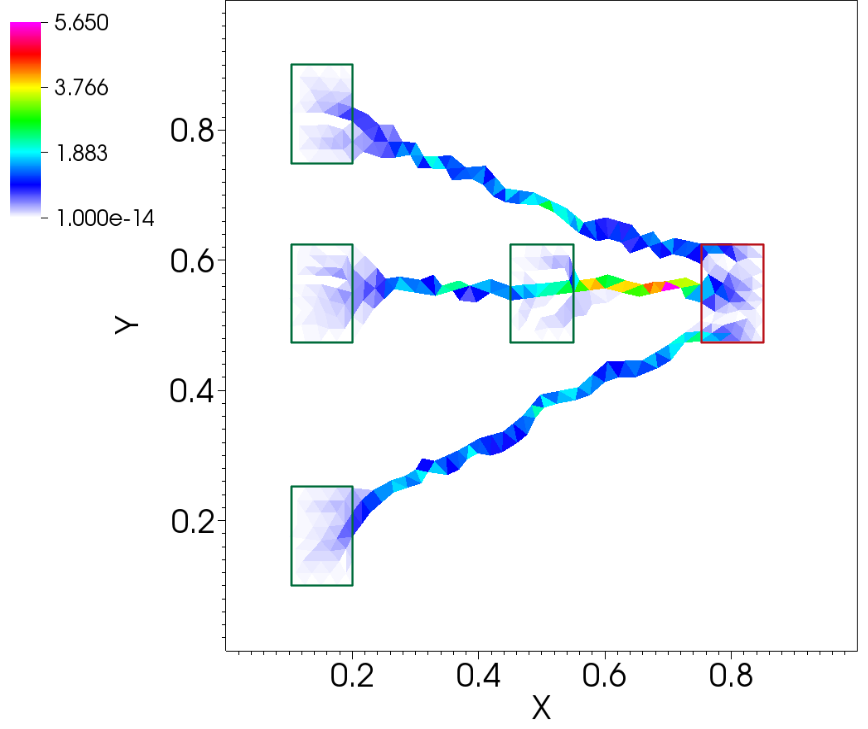} 
\caption{$\beta = 1.2$}
\label{fig:subim3}
\end{subfigure}
\end{figure}

In this work, solving the routing optimization problem consists of finding the steady state solution $(\mu^*, u^*):\Omega\rightarrow\mathbb{R}_{\geq0}\times \mathbb{R}$ of Eq. (\ref{eqn:ddmk}), i.e. $(\mu^*(x),u^*(x))=\lim_{t\to +\infty}(\mu(t,x),u(t,x))$. Numerical solution of the above model can be obtained by means of a double discretization in time and space\cite{facca2016towards,facca2019numerics,facca2020branch}. The resulting solver (called from now on \dmkS) has been shown to be efficient, robust and capable of identifying the typically singular structures that arise from the original problem. In Fig.~\ref{fig:beta}, some visual examples of the numerical $\mu^*$ obtained for different values of $\beta$ are shown. The same authors showed that the \dmkS\text{} is able to emulate the results for the discrete formulation of the PP model proposed by Tero et al.\cite{tero2007mathematical}
Under appropriate regularity assumptions, it can be shown~\cite{facca2019numerics,facca2020branch} that the equilibrium solution of the above problem $(\mu^*(x),u^*(x))$ is a minimizer of the following functional:
\bea\label{eqn:lyap}
\mathcal{L}(\mu,u)=\frac{1}{2}\int_{\Omega}\mu|\nabla u|^2 dx + \int_{\Omega}\frac{\mu^{P(\beta)}}{P(\beta)}dx \,,
\eea
where $P(\beta)=(2-\beta)/\beta$. In words, this functional is the sum of the total energy dissipated during transport (the first term is the Dirichlet energy corresponding to the solution of the first PDE) plus a nonlinear (sub-additive) function of the total capacity of the system at equilibrium. In terms of costs, this functional can be interpreted as the cost of transport, assumed to be proportional to the total dissipated energy, and the cost of building the transport infrastructure, assumed to be a nonlinear function (with power $2-\beta$) of the total transport capacity of the system.

%
%
We exploit the robustness of this numerical solver to extract the solutions of \dmk\text{} equations corresponding to various routing optimization problems. We here focus on the case $\beta\geq1$, where the approximate support of $\mu^*$ displays a network-like structure. This is the first step of our extraction pipeline, which we denote as \dmkS. The numerical solution of these equations does not allow for a straightforward network representation. Indeed, depending on various numerical details related to the spatial discretization and other parameters, one usually obtains a visually well-defined network structure (see Fig.~\ref{fig:subim1}) whose rendering as a graph object is however uncertain and non-unique. This in turns can hinder a proper investigation of the topological properties associated to optimal routing optimization problems, motivating the main contribution of our work: the proposal of a graph extraction pipeline to automatically and robustly extract network topologies from the solutions' output of \dmkS. We reinforce that our contribution is not limited to this application, but is also able to extract network-like shapes from any kind of image where a color or greyscale thresholds can be used to identify the sought structure.

Our extraction pipeline then proceeds with two main steps: \textit{pre-extraction} and \textit{graph filtering}. The first one tackles the problem of translating a solution from the continuous scenario into a graph structure, while the second one addresses the problem of removing redundant graph structure resulting from the previous step.  
A pseudo-code of the overall pipeline is provided in Algorithm \ref{alg:pipeline}\cite{mesh}.

\begin{algorithm}[htb]
   \caption{Generating optimal networks from solutions of dynamical systems: the pipeline}
   \label{alg:pipeline}
\begin{algorithmic}
  \STATE {\bfseries Input:} parameters to set up network extraction problem: \\
  \quad i) Set the space $\Omega$: $\ccup{T_{i}}_{i}$ grid triangulation and mesh-related parameters.\\
  \quad ii) Set up routing optimization problem: $\beta\geq1$, $ \mu_0,\,\  f^+,\ f^-$ and threshold $\delta$.\\
   \quad iii) Set up the \textit{discrete} routing optimization (for \textit{graph filtering}): $\beta_{d}, \delta_{d}$ and $\tau_{BC}$.
   \\
    \STATE {\bfseries Output:} $G(V,E,W)$ final network\\
   \STATE 1. \quad \textit{Run} \dmkS: outputs $(\mu^*, \ u^*)$ \\
   \STATE 2. \quad \textit{Graph pre-extraction}: outputs  $G(V,E,W)$ with possible redundancies \\
   \STATE 3. \quad \textit{Graph filtering}: removes redundancies from $G(V,E,W)$ \\
  \end{algorithmic}
\end{algorithm}

Our final goal is to translate the solution pair $(\mu^*, u^*)$ into a proper network structure using several techniques from graph theory. With these networks at hand, a practitioner is then able to investigate topologies associated with this novel representation of routing optimization solutions.




\section{Graph preliminary extraction}
\label{sec:graph_extraction}
In this section, we expand on the \textit{graph pre-extraction} step: extracting a network representation from the numerical solution output of the \dmkS. This involves a combination of numerical methods for discretizing the space and translating the values of $\mu^{*}$, and $u^{*}$ into edge weights of an auxiliary network, which we denote as $G=(V,E,W)$, where $V$ is the set of nodes, $E$ the set of edges and $W$ the set of weights.

The \dmk\text{} solver outputs the solution on a triangulation of the domain $\Omega$ (here also named \textit{grid}) and denoted as $\Delta_\Omega=\{T_i\}_i$, with $\cup T_{i}=\Omega$. The numerical solution, piecewise constant on each triangle $T_{i}$, is considered assigned to the triangle barycenter (center of gravity) at position $\v{b}_{i}=(x_{i},y_{i})\in \Omega$\cite{2d}. This means that the result is a set of pairs
$\ccup{\bup{ \mu^{*}(\v{b}_{i}),u^{*}(\v{b}_{i}) }}_{i}$.
We can track any function of these two quantities. For simplicity, we use $\mu^{*}$ (see Fig. \ref{fig:beta} for various examples), but one could use $u^{*}$ or a function of these two. This choice does not  affect the procedure, although the resulting network might be different. \\
We neglect information on the triangles where the solution is smaller than a user-specified threshold $\delta \in \mathbb{R}_{\geq0}$, in order to work only with the most relevant information. Formally, we only keep the information on $T_{i}$ such that $\mu^{*}(\v{b}_{i})\geq\delta $. We observed empirically that in many cases, several triangles contain a value of $\mu^{*}$ that is orders of magnitude smaller than others, see for instance the scale of Fig. \ref{fig:beta}.
Since we want to build a network that connects these barycenters, we remark that this procedure depends on the choice of the threshold $\delta$: if $\delta_1<\delta_2,$ then $G({\delta_2}) \subset G({\delta_1})$. On one hand, the smaller $\delta$, the more likely $G$ is to be connected, but at the cost of containing many possibly loop-forming edges and nodes (the extreme case $\delta=0$ uses the whole grid to build the final network); on the other hand, the higher $\delta$, the smaller the final network is (both in terms of the number of nodes and edges). Thus one needs to tune the parameter $\delta$ such that resulting paths from sources to sinks are connected while avoiding the inclusion of redundant information.

The set of relevant triangles does not correspond to a straightforward meaningful network structure, i.e. a set of nodes and edges connecting neighboring nodes. In fact, we want to remove as much as possible the biases introduced by the underlying triangulation and thus we start by connecting the triangle barycenters. For this, we need rules for defining nodes, edges and weights on the edges. Here, we propose three methods for defining the graph nodes and edges and two functions to assign the weights.
The overall \textit{graph pre-extraction} routine is given by choosing one of the former and one of the latter, and it can be applied also to more general inputs beyond solutions of the \dmkS.

\subsection{Rules for selecting nodes and edges}\label{subsec:grules}
Selecting $V$ and $E$ requires defining the neighborhood $\sigma(T_{i})$ of a triangle in the original triangulation $\Delta_{\Omega}$ (for $i$ such that $\mu^{*}(\v{b}_{i})\geq\delta$). We consider three different procedures:
\begin{enumerate}[label=(\Roman*)]
\item  \label{itm:ENsh} Edge-or-node sharing: $\sigma(T_{i})$ is the set of triangles that either share a grid edge or a grid node with $T_i$. 
\item  \label{itm:Esh}  Edge-only sharing: $\sigma(T_{i})$ is the set of triangles that share a grid edge with $T_i$. Note that $|\s(T_i)|\leq 3, \ \ \forall i$.
\item \label{itm:origT}Original triangulation: let $v,w,s$ be the grid nodes of $T_i$ ; then add $v,w,s$ to $V$ and $(v,w), (w,s),(s,v)$ to $E$.
\end{enumerate}

\begin{figure}[hptb]
	\centering
    \begin{subfigure}[b]{\textwidth}
        \includegraphics[width=\textwidth]{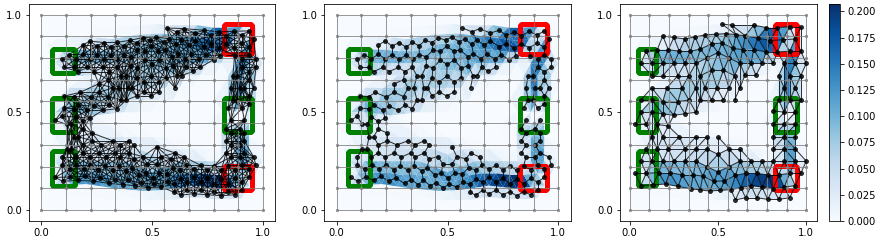}
    \end{subfigure} \label{graph_rep_1}
      \caption{\textit{Graph pre-extraction} rules. Left): edge-or-node sharing \ref{itm:ENsh}; center): edge-only sharing \ref{itm:Esh}; right): original triangulation \ref{itm:origT}. We monitor the conductivity $\mu$ and use parameters $\mu_0=1,\, \beta = 1.02, \, \delta=0.0001$. Weights $w_{ij}$ are chosen with $\text{AVG}$ \ref{itm:avg}, $f$ is chosen such that sources and sinks are inside green and red rectangles respectively. }
      \label{fig:Gextraction}
\end{figure}

 It is worth mentioning that since the grid $\Delta_\Omega$  is non uniform and $\mu^{*}$ is not constant, we cannot control \textit{a priori} the degree $d_{i}$ of a node $i$ in the graph $G$ generated for a particular threshold $\delta$.  
We give examples of networks resulting from these three definitions in Fig. \ref{fig:Gextraction} and a pseudo-code for the first two in Algorithm \ref{alg:ext}.

\begin{algorithm}[htb]
   \caption{Graph extraction}
   \label{alg:ext}
\begin{algorithmic}
  \STATE {\bfseries Input:} $\bup{\mu,u}$ solution of the \dmkS, 
   $\delta$ threshold, $\ccup{T_{i}}_{i}$ grid triangulation
    \STATE {\bfseries Output:} a network $G(V,E,W)$ \\ 
    \STATE { Initialize: $V,E=\emptyset$}
    \FOR {$i \;\; s.t. \;\;  \mu(\v{b}_{i})\geq\delta $}
    \STATE{ $V \leftarrow V \cup \ccup{i}$}
    \FOR { $T_{j} \in \sigma(T_i) \;\; s.t. \;\;  \mu(\v{b}_{j})\geq\delta $}
		\STATE{ $V \leftarrow V \cup \ccup{j}$ \\
		$E \leftarrow E \cup e_{ij}:=(i,j) $  \\
		$w_{ij}=f\bup{\mu(\v{b}_{i}),\mu(\v{b}_{j})}$  }
		\ENDFOR
	\ENDFOR
\end{algorithmic}
\end{algorithm}

\subsection{Rules for selecting weights}\label{subsec:wrules}
The weights $w_{ij}$ to be assigned to edges $e_{ij}:=(i,j) \in E$ should be a function of $\mu(\v{b}_{i})$ and $\mu(\v{b}_{j})$, the density contained in the original triangles. We consider two possibilities for this function:

\begin{enumerate}[label=(\roman*),ref=(\roman*)]
\item\label{itm:avg}  Average (AVG): $w_{ij}= \f{\mu(\v{b}_{i})+\mu(\v{b}_{j})}{2}$ .
\item\label{itm:er}Effective reweighing (ER): $w_{ij}= \f{\mu(\v{b}_{i})}{d_{i}}+\f{\mu(\v{b}_{j})}{d_{j}}$ .
\end{enumerate}
While using the average as in \ref{itm:avg} captures the intuition, it may overestimate the contribution of a triangle when this has more than one neighbor in $G$  with the risk of calculating a total density larger than the original output of the \dmkS. To avoid this issue, we consider an \textit{effective} reweighing as in \ref{itm:er}, where each triangle contribution by the degree $d_{i}=|\sigma_{i}|$ of a node $i\in V$ is reweighted, with $\sigma_{i}$ the set of neighbors of $i$. This guarantees the recovery of the density obtained from \dmkS, since $\f{1}{2} \sum_{i,j } w_{ij}= \f{1}{2}\sum_{i }\rup{\mu(\v{b}_{i}) + \sum_{j \in \sigma_{i}}\f{\mu(\v{b}_{j})}{d_{j}}}=\sum_{i} \mu(\v{b}_{i})$, where in the sum we neglected isolated nodes, i.e. $i$ s.t. $d_{i}=0$. Note that in the case of choosing the original triangulation for node and edge selection (case \ref{itm:origT} above), the ER rule does not apply; in that case, we use AVG. 

\section{Graph filtering}
\label{sec:graph_filtering}
The output of the graph extraction step is a network closer to our expectation of obtaining an optimal network topology resulting from a routing optimization problem. However, this network may contain redundant structures like dangling nodes or small irrelevant loops (see Fig.~\ref{fig:Gextraction}). These are not related to any intrinsic property of optimality, but rather are a feature of the discretization procedure resulting from the graph pre-extraction step.  It is thus important to filter the network by removing these redundant parts. However, how to perform this removal in an automated and principled way is not an obvious task. One has to be careful in removing enough structure, while not compromising the core optimality properties of the network. This removal is then a problem in-and-of-itself,  we name it \textit{graph filtering} step.
We now proceed by explaining how we tackle it in a principled way and discuss its quantitative interpretation in terms of minimizing a cost function interpolating between an operating and an infrastructural cost. 

\subsection{The \textit{Discrete} \dmkS}\label{subsec:dmkdiscrete}
Going beyond heuristics and inspired by the problem presented in Sec. \ref{sec:routingopt}, we consider as a solution for the graph filtering step, the implementation of a second routing optimization algorithm to the network $G$ output of the pre-extraction step, i.e. in discrete space. 
Several choices for this could be drawn, for instance, 
from routing optimization literature \cite{oliveira2011mathematical}, but we need to make sure that this second optimization step does not modify any of the intrinsic properties related to optimality resulting from the \dmkS. 
We thus propose to use a \textit{discrete} version of the \dmkS \text{ }(\ddmk). This was proven to be related to the \textit{Basis Pursuit} (BP) optimization problem \cite{facca2018physarum}. In fact, BP is related\cite{straszak2016irls} to the PP dynamical problem in discrete space and the \ddmk \text{} gives a solution to the PP in discrete space\cite{facca2018physarum}. The discretization results in a reduction of the computational costs for solutions of BP problems, compared to standard combinatorial optimization approaches\cite{facca2018physarum}. Being an adaptation to discrete settings of our original optimization problem, it is a natural candidate for a graph filtering step, preserving the solution's properties.

\begin{figure}[ht]
    \centering 
        \includegraphics[width=\textwidth]{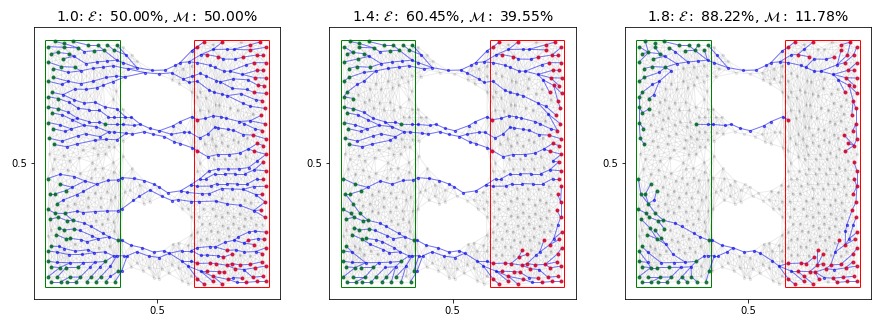}   
        \caption{\textit{Graph filtering} rules. Left):
            $\beta_{d}=1.0$; center): $\beta_{d}=1.4$;
            right): $\beta_{d}=1.8$. The number on top
            denote the contributions of operating and
            infrastructural cost to the energy. Green and red dots represent sources and sinks respectively ($\tau_{BC}=10^{-1}$); blue edges are those $e$ with $\mu_e^*\geq 10^{-3}$. The filtering input is generated from \dmkS \text{} with $\beta=1.05$. The apparent lack of symmetry of the network's branches is due to numerical discretization of the domain, solver and threshold $\delta$. As the relative size of the terminal set decreases compared to the size of remaining part of the domain, this lack of symmetry becomes negligible.} 
      \label{fig:graph_filtering}
\end{figure}

The problem is stated as follows. Consider the \textit{signed incidence} matrix $\mathbf{B} \in \mathbb{M}_{N\times M}$ of a weighted graph $G=(V,E,W)$, with entries $B_{ie}=\pm 1$ if the edge $e$ has node $i$ as start/end point, 0 otherwise; $N=|V|$ and $M=|E|$. 
Denote $\v{\ell} = \ccup{\ell_{e}}_{e}$ the vector of edge lengths,  $\mathbf{f}$ a $N$-dim vector of source-sink
values with entries satisfying $\sum_{i\in V}f_{i}=0$; this is the discrete analogues of the source-sink function $f(x)$ introduced in Section \ref{sec:routingopt}; the functions $\mu(t)\in \mathbb{R}^{M}$ and $u(t)\in \mathbb{R}^{N}$ correspond to the \textit{conductivity} and \textit{potential} respectively, similarly to the continuous case, but this time they are vectors with entries $\mu_{e}(t)$ and $u_{i}(t)$ defined on edges and nodes respectively.
The PP discrete dynamics corresponding to the original routing optimization problem can be written as: 
\bea\label{eqn:discreteDMK1}
 f_{i}  &=&\sum_{e} B_{ie} \f{\mu_{e}(t)}{\ell_{e}} \sum_{j}B_{ej} \, u_{j}(t) \,, \label{eqn:kirkdiscrete}\\ 
\mu_{e}'(t)&=&\rup{\f{\mu_{e}(t)}{\ell_{e}}|\sum_{j}B_{ej}\,u_{j}(t)|}^{\beta_{d}}-\mu_{e}(t)  \label{eqn:mudiscrete}\,,\\ 
\mu_{e} (0)&>&0 \,,\label{eqn:ICdiscrete}
\eea
where 
$|\cdot|$ is the absolute value element-wise. Equation (\ref{eqn:kirkdiscrete}) corresponds to Kirchoff's law, Eq. (\ref{eqn:mudiscrete}) is the discrete dynamics with $\beta_{d}$ a parameter controlling for different routing optimization mechanisms (analogously to $\beta$ in Eq. \ref{eqn:ddmk2});  Eq. (\ref{eqn:ICdiscrete}) is the initial condition.
The importance of this systems stems in having an interesting theoretical correspondence: its equilibrium point corresponds to the minimizer of a cost function analogous to Eq.~(\ref{eqn:lyap}) that, similarly to the continuous case, can be interpreted as global energy functional. This is:
\be
\mathcal{L}_{\beta}(\mu(t))= \f{1}{2} \sum_{e}\mu_{e}(t)\bup{\f{1}{\ell_{e}}\sum_{j}B_{ej}\, u_j(\mu(t))}^{2}\ell_{e} +\f{1}{2}\sum_{e}\f{\mu_{e}(t)^{P(\beta)}}{P(\beta)}\,\ell_{e} \,, 
\ee
where $P(\beta)={2-\beta}/{\beta}$ and $u(\mu(t))$ is a function implicitly defined as the solution of Eq. (\ref{eqn:kirkdiscrete}). The first term corresponds to the energy dissipated during transport, it can be interpreted as the operating costs, whereas the second is the infrastructural cost. The equilibrium point of $\mu_{e}(t)$ is stationary at the previous energy function, and for $\beta_{d}=1$ it acts also as the global minimizer due to its convexity. For $\beta_{d}>1$ the energy is not convex, thus in general the functional will present several local minima towards which the dynamics will be attracted. The case $\beta_{d}<1$ does not act as a filter because it encourages trajectories to spread through the network, instead of removing edges, and so not interesting to our purposes. Discretization in time of Eq. (\ref{eqn:mudiscrete}) by the implicit Euler scheme combined with Newton method leads to an efficient numerical solver, see Facca et al. \cite{facca2018physarum} for more details. The above scheme gives the solution to the BP problem and represents the \ddmk.
Similarly to the graph pre-extraction step, the filtering  is also valid beyond networks related to solutions of the \dmkS. It applies to more general inputs if defined on a discrete space, for instance, images. Finally, notice that the filter generates a graph with a new set of nodes and edges, both subsets of the corresponding ones in $G$, result of the pre-extraction. The weights of the final graph can then be assigned with same rules as in \ref{subsec:grules}; in addition, one can consider as weights the values of $\mu_{e}$ resulting from the BP problem (we named this weighing method ``BPW''). Alternatively, one can ignore the weights of BP and keep (for the edges remaining after the filter) the weights as in the previous pre-extraction step (labeled as ``IBP''). Figure \ref{fig:graph_filtering} shows an example of three filtering settings on the same input. 

\subsection{Selecting sources and sinks}
The \ddmk\text{} requires in input a set of source and sink nodes ($S^{+}$ and $S^{-}$)  that identify the support of the forcing vector $\v{f}$ introduced in \ref{subsec:dmkdiscrete}. However, the graph pre-extraction output $G$  might contain redundant nodes (or edges) as mentioned before. In principle, among the nodes $i \in V$, all of those contained in the support of $f(x=\v{b}_{i})$, i.e. contained in the supports of sources and sinks of the original routing optimization problem in Eq.~(\ref{eqn:ddmk1}), are \textit{eligible} to be treated as sources or sink in the resulting network. However, several paths connecting source and sink nodes may be redundant and clearly not compatible with an optimal routing network (see Supplementary Fig.~S2 for such an example). Therefore, it is important to select ``representatives'' for sources and sinks, such that the final network is heuristically closer to optimality. Here we propose a criterion to select source and sink nodes from the eligible ones in each of the connected components $\{C_m\}_m$ of $G$, using a combination of two network properties. 
Starting from the complete graph formed by all the nodes characterized by a significant (above the threshold) density, source and sink nodes and rates are defined as follows.
A node $i \in S^{+}$, i.e. is a source $f_{i}>0$, if either i) is in the convex hull of the set of eligible sources or ii) its betweenness centrality is smaller than a given threshold $\tau_{BC}$. Similarly for sink nodes in $S^{-}$. 
This is because, on one side, nodes in the convex hull capture the outer shape structure of the source and sink sets defined in the continuous problem;  on the other side, nodes with \textit{small} values of the betweenness centrality capture the end-points of $G$ inside the source and sink sets, analogously to leaves (i.e., degree-one nodes)\cite{connectivity}. We present these ideas in more detail in the Supplementary Fig. S2.
Once we have identified the sets of source and sink vertices, we need to assign a proper value $f_{i}$ such that Kirchhoff law is satisfied in each of the different connected components $C_{m}$. It is reasonable to assume that each connected component is ``closed'', i.e. $\sum_{i \in C_{m}} f_{i}=0 \,, \forall C_{m}$. Denoting with $|S|$ the number of elements in a set $S$ and $V(C_m)$ the set of nodes in $C_{m}$, we then distribute the mass-fluxes uniformly by setting  $f_{i}=\f{1}{|S^{+}\cap V(C_m)|}$ for $i \in S^{+}$, and $f_{i}=-\f{1}{|S^{-}\cap V(C_m)|}$ for $i \in S^{-}$ sinks ($f_{i}=0$ otherwise) so that the total original source and sink flux is assigned to the overall source/sink nodes of all $C_{m}$. 
Note that this procedure maintains the overall system and each connected component ``closed'', as stated above.




\section*{Model validation}
\label{sec:validation}
Our extraction pipeline proceeds by compressing routing information in the raw output of the \dmkS \text{} (although what follows is not restricted to this case) on a lean network structure. This might lead us to lose relevant information in the process. Hence, we need to devise \textit{a posteriori}  estimates that provide quantitative guidance on the ``leanness'' and information loss of the final network. Here we propose metrics to evaluate the compression performance of the various graph pre-extraction and filtering protocols. The raw information is made of a set of weights $w(T_{i})$ representing the values $(\mu^{*},u^{*})$ on each of the triangles $T_{i}\in \Omega$. We consider as the truth benchmark the distribution of $w$, or any other quantity of interest, supported on the subgrid $\Delta^{\delta}_{\Omega}\subset\Delta_{\Omega}$ formed by all triangles where $w$ is larger than the threshold value $\delta$, i.e., $\Delta_\Omega^{\delta}:=\{T_i \in \Delta_\Omega: w(T_i)\geq \delta\}$. We expect that a good compression scheme should preserve both the total \textit{amount} of the weights from the original solution in $\Delta^{\delta}_{\Omega}$ and the information of \textit{where} these weights are located inside the domain $\Omega$. Also, we want this compression to be parsimonious, i.e. to store the least amount of information as possible. We test against these two requirements by proposing two metrics that measure: i) an information difference between the raw output of the \dmkS \text{} and the network extracted using our procedure, capturing the information of where the weights are located in space; ii) the amount of information needed to store the network. 
 
\begin{figure}[hptb]
    \centering
    \begin{subfigure}[a]{0.49\textwidth}\label{fig:Wmeasure}
        \includegraphics[width=\textwidth]{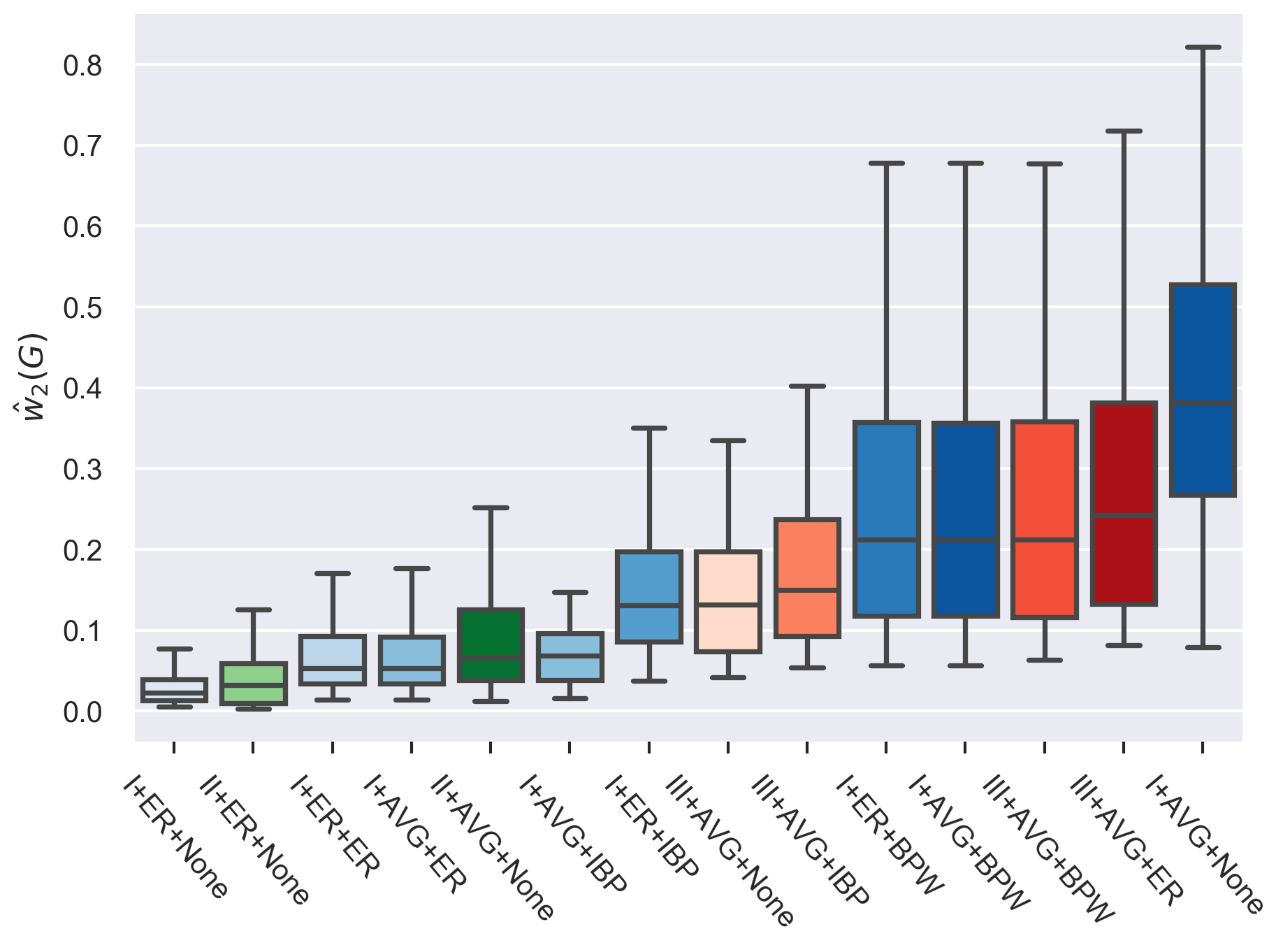}
    \end{subfigure}
        \begin{subfigure}[a]{0.49\textwidth}\label{fig:Lmeasure}  
        \includegraphics[width=\textwidth]{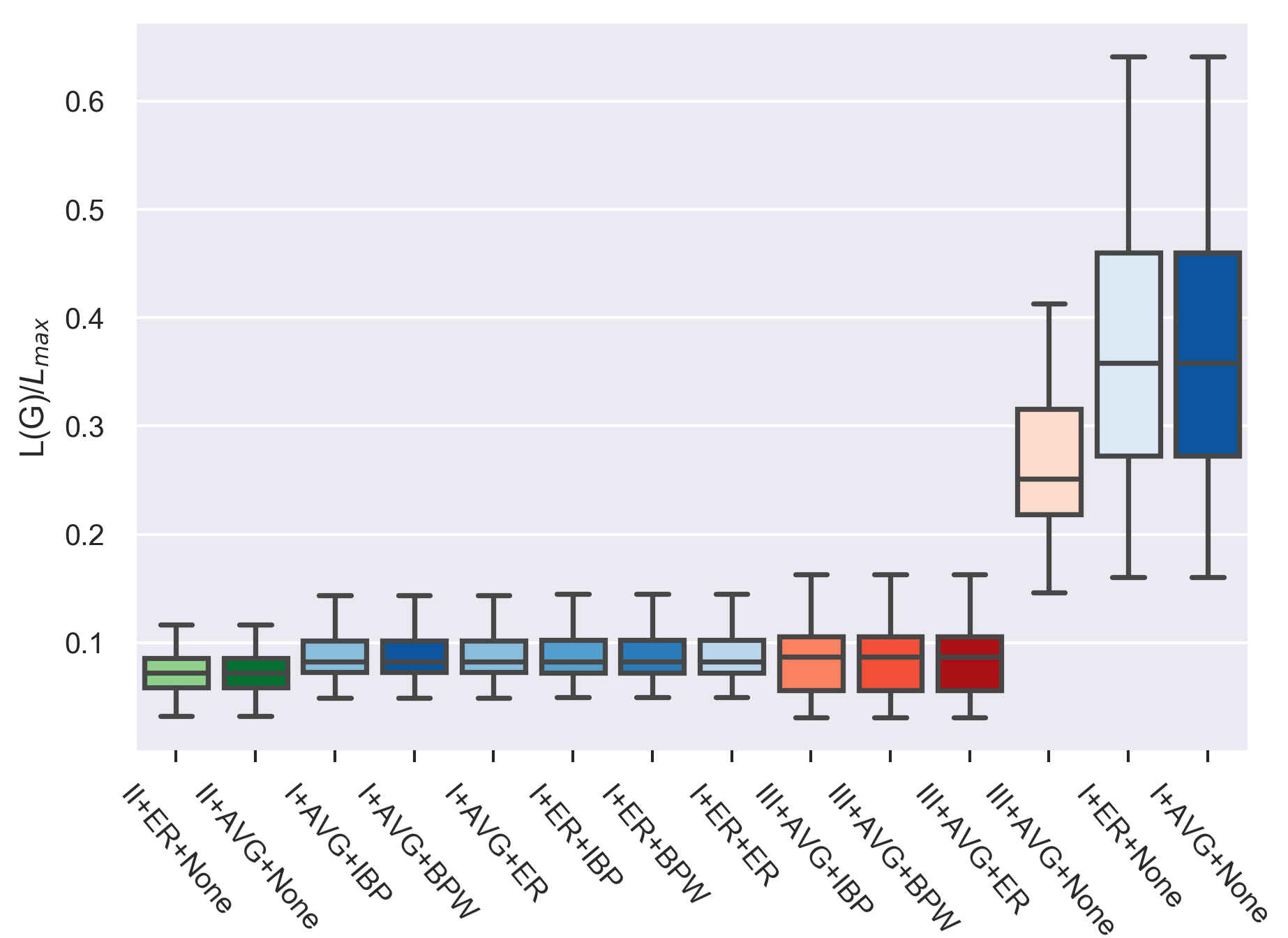}
    \end{subfigure}
      \caption{Graph extraction performance evaluation. We plot results for \nix{the} different combinations of the graph extraction rule in terms of: (left) the  metric $\hat{w}_{q}(G)$ of Eq. (\ref{eqn:metric}); (right) total network length $L(G)$ normalized by $L_{max}$, the max length over the 170 networks. Each bar denotes a possible combination as follows: roman numbers denote one of the three rules I-III (\ref{subsec:grules}); first label after the number denotes one of the rules to assign weights i-ii \ref{subsec:wrules}), which is applied to the output of the first step; the second (and last) label denotes the same rule but applied after the filtering step, ``None'' means that nothing is done, i.e. no filter applied, ``IBP'' means filter applied but with no reweighing, i.e. when an edge is removed by the filter we simply lose information without relocating its weight. Bars are color-coded so that  rules I-III have three different primary colors and their corresponding routines have different shades of that main color. Here, we keep track of the conductivity $\mu$ and show medians and quartiles of a distribution over 170 networks generated with $\beta \in \ccup{1.1,\, 1.2,\, 1.3}$, $\beta_{d}=1.1$ and $\delta=0.01$.
}
      \label{fig:evaluation}
\end{figure}

 Our first proposed metric relies in partitioning $\Omega$ in several subsets and  then calculating the difference in the extracted network weights and the uncompressed output, \textit{locally} within each subset. 
More precisely, we partition $\Omega$ into $P$ non intersecting subsets $C_{\alpha} \subset \Omega$, with $\alpha=1,\dots,P$ and $\cup_{1}^{P}C_{\alpha}=\Omega$. For example,  we define $C_\alpha = [x_i,x_{i+1}]\times [y_j,y_{j+1}]$, for $x_i,x_{i+1},y_j$ and $y_{j+1},$ consecutive elements of $N$-regular partitions of $[0,1]$, and $P=(N-1)^2$. Denote with $w_{\delta}(T_{i})$ the weight on the triangle $T_{i}\in\Delta^{\delta}_{\Omega}$, resulting from the \dmkS \text{}  (usually a function of $\mu^*$ and  $u^*$). If we denote the \textit{local weight} of $\Delta_\Omega^{\delta}$ inside $C_{\alpha}$ as $w_{\alpha}=\sum_{i: \v{b}_{i}\in C_{\alpha}}w_{\delta}(T_{i})$, 
 then we propose the following evaluation metric:
\bea\label{eqn:metric}
\hat{w}_q(G) &:=& \dfrac{1}{P}\rup{\sum_{\alpha=1}^{P}\left( \sum_{e\in E}|\, \mathbb{I}_\alpha(e) \ w_{e} - w_{\alpha}\,|\right)^{q} }^\frac{1}{q} \,,
\eea
where $\mathbb{I}_\alpha(e)$ is an indicator of whether an edge $e=(i,j) \in E$ is inside an element $C_{\alpha}$ of the partition, i.e. $\mathbb{I}_\alpha(e)=1,0,1/2$ if both $\v{b}_{i},\v{b}_{j}$ are in $C_{\alpha}$, none of them are, or only one of them is, respectively.
In words, $\hat{w}_{q}(G)$ is a distance between the weights of the network extracted by our procedure and the original weights output of the \dmkS, over each of the local subsets $C_{\alpha}$. This metric penalizes networks that either place large-weight edges where they were not present in the original triangulation, or low-weight ones where they were instead present originally. In this work we consider the Euclidean distance, i.e. $q=2$, but other choices are also possible. 
Note that $\hat{w}_{q}(G)$ does not say anything about how much information was required to store the processed network. If we want to encourage parsimonious networks, i.e. networks with few redundant structures, then we should include in the evaluation the monitoring of $L(G)=\sum_{e\in E}  \ell_{e}$, the total path length of the compressed network, where the edge length $\ell_{e}$ can be specified based on the application. Standard choices are uniform $\ell_{e}=1,\, \forall e$ or the Euclidean distance between $\v{b}_{i}$ and $\v{b}_{j}$. Intuitively, networks  with small values of both $\hat{w}_{q}(G)$ and $L(G)$ are both accurate and parsimonious representations of the original \dmk \text{} solutions defined on the triangulation.\\ 
We evaluate numerous graph extraction pipelines in terms of these two metrics on various routing optimization problem settings and parameters. 
In Fig. \ref{fig:evaluation} we show the main results for a distribution of 170 networks obtained with $\beta \in \ccup{1.1,\, 1.2,\, 1.3}$ and $\beta_{d}=1.1$. Similar results were obtained for other parameter settings. Networks are generated as follows: first, we choose a set of $5$ different initial transport densities $\mu_0$, grouped in parabola-like, delta-like and uniform distributions, and a set of $12$ different configurations for sources/sinks (mainly rectangles placed in different positions along the domain, see Supplementary Information for more details). Then, for each of these setup, we run our procedure: i) first the \dmkS \text{} calculates the solution of the continuous problems; ii) then we apply the \textit{graph pre-extraction} procedure according \nix{to the} rules of Sec. \ref{subsec:grules} and weights as in Sec. \ref{subsec:wrules}; iii) finally, we run the \textit{graph filtering} step and consider various weight functions, as described in Fig. \ref{fig:evaluation}.

\begin{figure}[hptb]
    \centering
    \begin{subfigure}[a]{0.42\textwidth}\label{fig:finalEx1}
        \includegraphics[width=\textwidth]{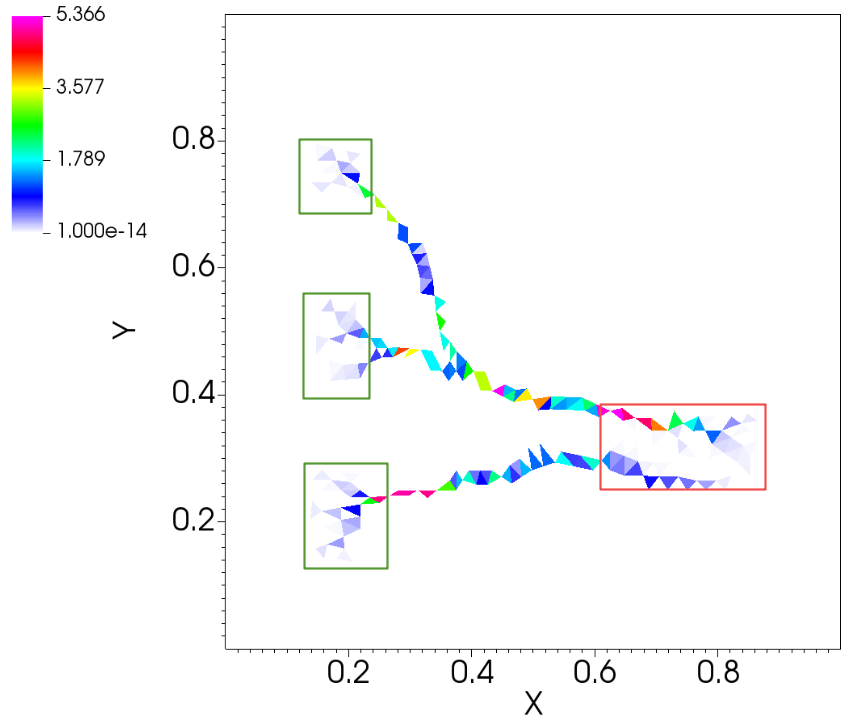}
    \end{subfigure}	
        \begin{subfigure}[a]{0.377\textwidth}\label{fig:finalEx2}  
        \includegraphics[width=\textwidth]{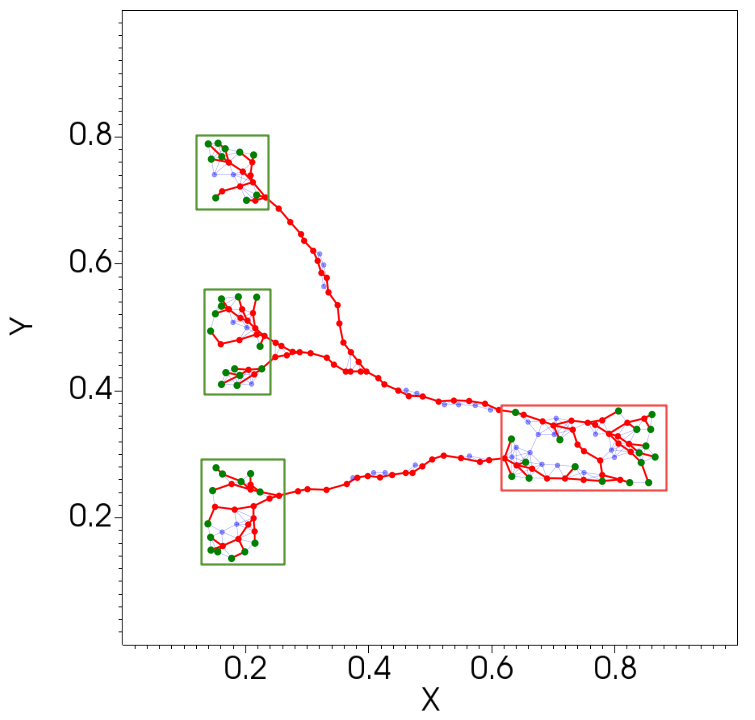}
    \end{subfigure}
      \caption{Network extraction example. We show a network generated from a routing optimization problem with parameters $\beta_c=1.4$, $\beta_d=1.3$ and $\delta=0.001$;  (left) raw output of the \dmkS; (right) final network extracted using the routine I-ER-ER.
}
      \label{fig:finalExample}
\end{figure}

We observe that not applying the final filtering step and considering rule I with ER to build the graph (I-ER-None), the values of $\hat{w}_2(G)$ are smaller than other cases. This is expected as by filtering we remove information and thus achieve better performance with this metric when compared to no filtering. However, we pay a price in terms of total relative length as $L(G)/L_{max}$ is larger for this case. When working with rule II, we notice the appearance of many non-optimal small disconnected components and this effect deteriorates if filtering is activated. Corresponding statistics show low values for both $\hat{w}_2(G)$ and $L(G)/L_{max}$. We argue that this is because rule II produces, by construction, fewer redundant objects than rule I in the initial phase. This might have a similar effect as a filter but is done \textit{a priori} during the pre-extraction, because rule II produces in this phase a limited number of effective neighbors. However, this comes at a price of higher variability with the sampled networks, as the variance of $\hat{w}_2(G)$ is higher than for the other combinations. Among the possibilities with filtering applied, we observe that rule I performs better than rule III, while all the weighting rules give a similar performance in terms of both metrics. Any combination involving rule I plus filtering has a similar performance as rule II in terms of both metrics but with smaller variability. Finally, these combinations perform differently in terms of the number of disconnected components (not shown here), with rule II producing more spurious splittings, as already mentioned. Depending on the application at hand, a practitioner should select one of these combinations based on their properties as discussed in this section. We give an example of a network generated with I-ER-ER in Fig. \ref{fig:finalExample}.  

\section*{Application: network analysis of a vein network}
\label{sec:qualitative}

\begin{figure}[hptb]
    \centering
    \begin{subfigure}[b]{0.42\textwidth}
        \includegraphics[width=\textwidth]{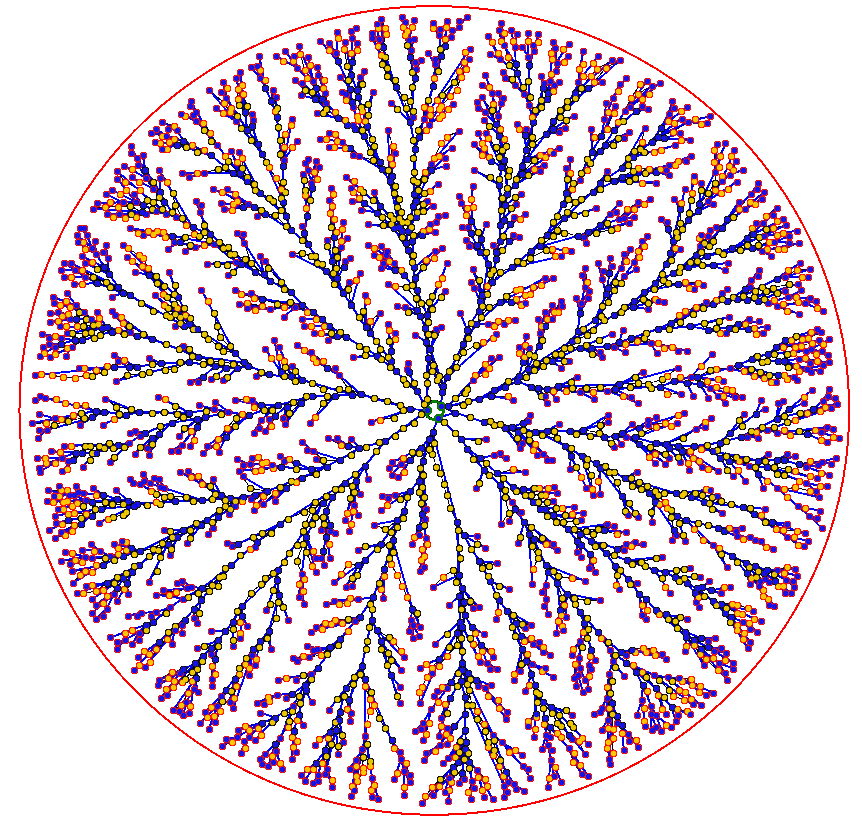}
    \end{subfigure} \label{fig:fungi}
        \begin{subfigure}[b]{0.42\textwidth}
        \includegraphics[width=\textwidth]{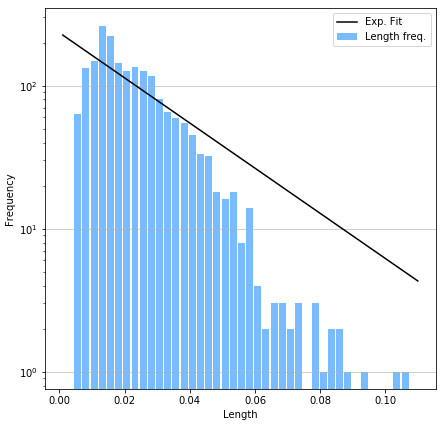}
    \end{subfigure} \label{fig:fungiL}  
    \caption{Application to fungi network. We generate
        a synthetic network similar to the image of Fig. 1a
        reported in Boddy et al. \cite{boddy2010fungal} and Fig. 4a in Obara et al.
        \cite{obara2012bioimage} for the of
        \textit{Phanerochaete velutina} fungus
        \cite{fricker2007network} and Fig. 1 in their
        supplementary for the
        \textit{Coprinus picaceus}. Fitted parameters are:
        $P_0=234.00, \gamma = 36.32$.  Here $f^+(x,y)=1,$ if $(x-0.5)^2+(y-0.5)^2\leq 0.01$; $\ f^+(x,y)=0,$ otherwise; $f^-(x,y)=-1,$ if $0.01<(x-0.5)^2+(y-0.5)^2\leq 0.45$; $f^-(x,y)=0$, otherwise. The network on the left corresponds to the filtered graph. Yellow nodes are degree-2 nodes that we omitted when computing the length distribution.  Green and red outlines are used to denote nodes in $S^+$ and $S^-$, respectively.}
      \label{fig:ppreal}
\end{figure}

\begin{figure}[hptb]
    \centering
    \begin{subfigure}[b]{0.42\textwidth}
        \includegraphics[width=\textwidth]{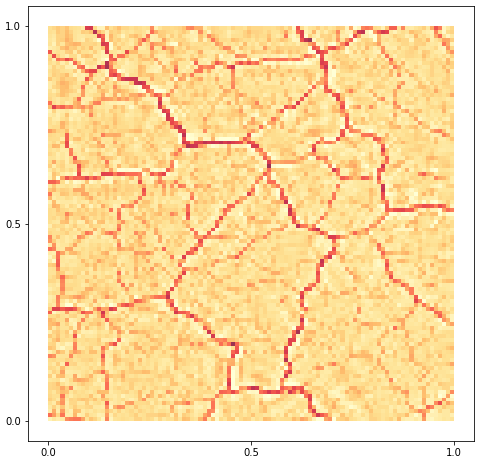}
    \end{subfigure} \label{fig:image1}
        \begin{subfigure}[b]{0.42\textwidth}
        \includegraphics[width=\textwidth]{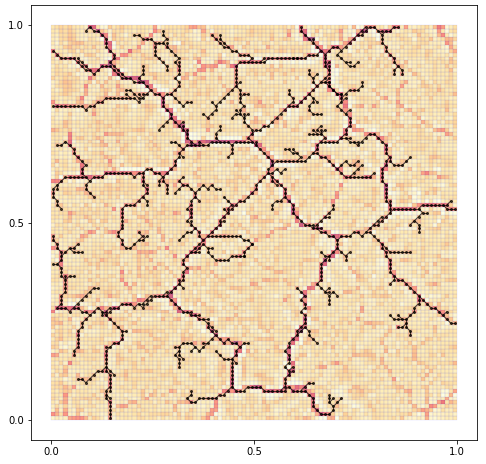}
    \end{subfigure} \label{fig:image2}  
      \caption{Application to images. We take an image of the \textit{P. Polycephalum} from the SMGR repository \cite{dirnberger2017introducing}.The picture used is a 1200x1200-pixel section of an original image of size 5184x3456 pixels (see Supplementary S4 for details) and extract a network with step 2 and 3 of our protocol. As a pre-processing step, we  downsample the image using \textit{OpenCV} (left) and use the color scale defined on the pixels as an artificial $\mu^*$ function. Using this information and the grid structure associated to the image's pixels, we first build a graph $G$ with the \textit{graph pre-extraction} step described in Sec. \ref{sec:graph_extraction}; then, we obtain a graph $G_{f}$ (right) using the \textit{graph filtering} step of Sec. \ref{sec:graph_filtering}, for an appropriate selection of sources and sinks, and adding a correction to retrieve loops. Notice that our protocol in its standard settings with filtering can only generate tree-like structures. Therefore, if we want to obtain a network with loops, we should consider a minor modification of our routine, which can be done in a fully automatized way, as explained in more details in the Supplementary S4.}
      \label{fig:imageEx}
\end{figure}

We demonstrate our protocol on a biological network of fungi foraging for resources in space. The network structure corresponds to the fungi response to food cues while foraging \cite{fricker2007network}. Edges are veins or venules and connect adjacent nodes. 
This and those of other types of fungi are well known networks typically studied using image segmentation methods \cite{baumgarten2010detection,obara2012bioimage,bebber2007biological,boddy2010fungal}. It is thus interesting to compare results found by these techniques and by our approach, under the conjecture that the underlying dynamic driving the network structure could be the same as the optimality principles guiding our extraction pipeline.
In particular, we are interested in analyzing the distribution $P(\ell)$ of the veins lengths, i.e. the network edges. The benchmark $P(\ell)$ distribution obtained by Baumgarten and Hauser\cite{baumgarten2010detection} using image processing techniques is an exponential of the type $P(\ell) = P_{0}\,e^{-\gamma \ell}$. Accordingly, as shown in  Fig. \ref{fig:ppreal}, we find that an exponential fit (with values $P_0=234.00, \gamma = 36.32$) well captures the left part of the distribution, i.e. short edges. Differences between fit and observed data can be seen in the right-most tail of the graph, corresponding to longer path lengths, where the data decay faster than the fit. However, we find that the exponential fit is nevertheless better than other distributions, such as the gamma and log-normal proposed in Dirnberger and Mehlhorn\cite{dirnberger2017characterizing} for the \textit{P. polycephalum}. Drawing definite quantitative conclusions is beyond the scope of our work, as this example aims at a qualitative illustration of possible applications that can be addressed with our model. In general, however, it seems not possible to choose a single distribution that well fits both center and tails of the distribution for various datasets of this type\cite{dirnberger2017characterizing}. \\
To conclude, we demonstrate the flexibility of our graph extraction method on a more general input than the one extracted from \dmkS. Specifically, we consider as example an image of \textit{P. polycephalum} taken from data publicly available in the Slime Mold Graph Repository (SMGR) repository \cite{dirnberger2017introducing}. We first downsample an image of the SMGR's \textit{KIST Europe data set}, using \textit{OpenCV} (left) and a color scale defined on the pixels as an artificial $\mu^*$ function. We build a graph using the \textit{graph pre-extraction} and \textit{graph filtering} steps as shown in Fig. \ref{fig:imageEx}. Notice that our protocol in its standard settings with filtering can only generate tree-like structures. Therefore, if we want to obtain a network with loops as we did in Fig. \ref{fig:imageEx}, we should consider a modification of our routine, which can be done in a fully automatized way, as explained in more details in the Supplementary S4. In short, after the \textit{graph pre-extraction} step, where loops are still present, we extract a tree-like structure close to the original loopy graph and give this in input to the filtering. We can then add \textit{a posteriori} edges that connect terminals that were close by in the graph obtained from the pre-extraction step but removed by the filter, thus recovering loops. In case obtaining loops is not required, our routine can be used with no modifications. Adapting our filtering model to allow for loopy structures in a principled way, analogously to what done in Sec. 3, will be subject of future work.

\section*{Discussion}

We propose a graph extraction method for processing raw solutions of routing optimization problems in continuous space into interpretable network topologies. The goal is to provide a valuable tool to help practitioners bridging the gap between abstract mathematical principles behind optimal transport theory and more interpretable and concrete principles of network theory. While the underlying routing optimization scheme behind the first step of our routine uses recent advances of optimal transport theory, our tool enables automatic graph extraction without requiring expert knowledge.
We purposely provide a flexible routine for graph extraction so that it can be easily adapted to serve the specific needs of practitioners from a wider interdisciplinary audience. We thus encourage users to choose the parameters and details of the subroutines to suitably customize the protocol based on the application of interest. To help guiding this choice, we provide several examples here and in the Supplementary Information.
We anticipate that this work will find applications beyond that of automating graph extraction from routing optimization problems. We remark that two of the three steps of our protocol apply to inputs that might not necessarily come from solutions of routing optimization. Indeed, the pipeline can be applied to any image setting where an underlying network needs to be extracted. The advantage of our setting with respect to more conventional machine learning methods is that the final structure extracted with our approach minimizes a clearly defined energy functional, that can be interpreted as the combination of the total dissipated energy during transport and the cost of building the transport infrastructure.
We foresee that this minimizing interpretation together with the simplification of the pipeline from abstract modeling to final concrete network outputs will foster cross-breeding between fields as our tool will inform network science with optimal transport principles and vice-versa. In addition, we expect to advance the field of network science by promoting the creation of new network databases related to routing optimization problems.


\bibliography{references}

\begin{thebibliography}{10}
\bibitem{hydraulic}
\bibinfo{title}{With hydraulic interpretation, we can think of the edges as
  pipes, small cylindrical channels where the mass is passing through, and the
  capacity is proportional to the size of the pipe diameter.}
\bibitem{2d}
\bibinfo{title}{In this work we focus on a 2d space, but the procedure can be
  generalized to 3d.}
\bibitem{connectivity}
\bibinfo{title}{Due to the high graph connectivity, degree centrality is not
  appropriate for selecting these ending parts.}
\end{thebibliography}

\section*{Acknowledgements}
The authors thank the International Max Planck Research School for Intelligent Systems (IMPRS-IS)
for supporting Diego Baptista and Daniela Leite.

\section*{Additional information}
\textbf{Accession codes}:  open source codes and executables are available at \url{https://github.com/Danielaleite/Nextrout}. \\

\clearpage



\newcommand{\beginsupplement}{%
        \setcounter{table}{0}
        \renewcommand{\thetable}{S\arabic{table}}%
        \setcounter{figure}{0}
        \renewcommand{\thefigure}{S\arabic{figure}}%
        \setcounter{equation}{0}
        \renewcommand{\theequation}{S\arabic{equation}}
         \setcounter{section}{0}
        \renewcommand{\thesection}{S\arabic{section}}
 }

\beginsupplement

\section*{{Supporting Information (SI)}}


\section{Examples of \textit{graph pre-extraction} routines}
\label{apx:pre-extraction}

We provide here several examples of networks resulting from different routing optimization setup for each of the three graph definitions. 

\begin{figure}[hptb]
    \centering
    \begin{subfigure}[a]{0.95\textwidth}\label{apxFig:beta15}
        \includegraphics[width=\textwidth]{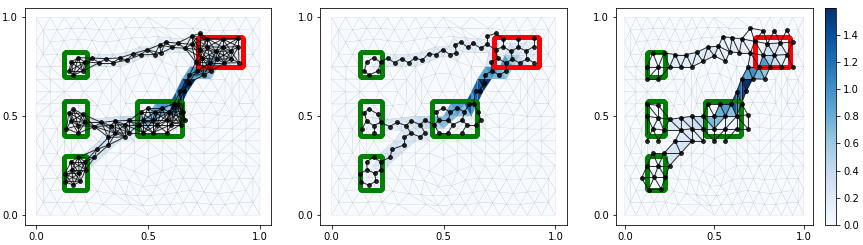}
           \caption{$\beta=1.1$: traffic consolidation}
      \end{subfigure}
       \begin{subfigure}[a]{0.95\textwidth}\label{apxFig:beta1}
        \includegraphics[width=\textwidth]{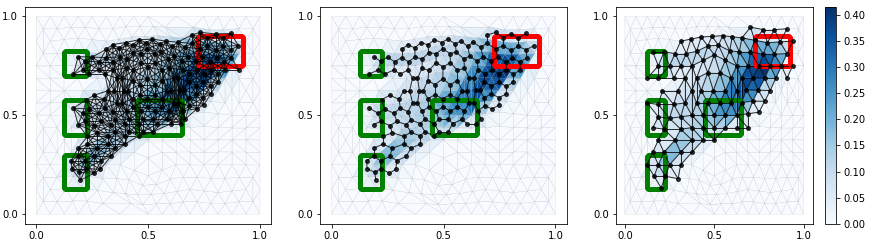}
        \caption{$\beta=1.0$: shortest-path optimization}
      \end{subfigure}
      \caption{\textit{Graph pre-extraction} rules for three transportation optimization problems. Left): edge-or-node sharing; center): edge-only sharing; right): original triangulation. We monitor the density $\mu$ and use parameters: $\mu_0=1$, $ f=\text{5rch}$, $\delta=0.0001$, $w_{ij}=\text{ER}$. }
    \label{apxFig:beta}
\end{figure}

\newpage
\section{Examples networks and routing optimization scenarios}
\label{apx:examples}

We provide here several examples of networks resulting from different routing optimization setups for several choices of our proposed routines. This should serve as an example guideline on what parameters to choose based on the application. More specifically, we show three different setups given in input as initial routing optimization problems to the \dmkS \text{}. We consider: i) different locations of sources and sinks to show how this impacts the formation of branches and their symmetry; ii) different values of $\beta \in \ccup{1.2, 1.3, 1.5}$ to show how traffic consolidates into fewer edges as $\beta$ grows; iii) different initial $\mu_{0}(x,y)$ (parabolic, delta-like and uniform) to show variability in the initial transport density. 

\begin{figure}[hptb]
    \centering
    \begin{subfigure}[a]{0.33\textwidth}\label{apxFig:step1}
        \includegraphics[width=\textwidth]{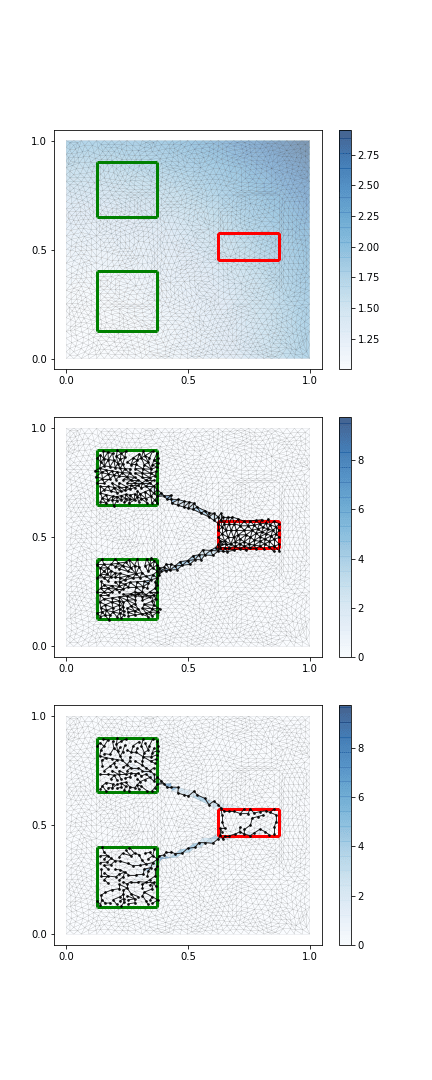}
        \vspace{-2cm}
           \caption{Example 1: parabolic transport density, $\beta = 1.2$}
      \end{subfigure}
       \begin{subfigure}[a]{0.33\textwidth}\label{apxFig:step2}
        \includegraphics[width=\textwidth]{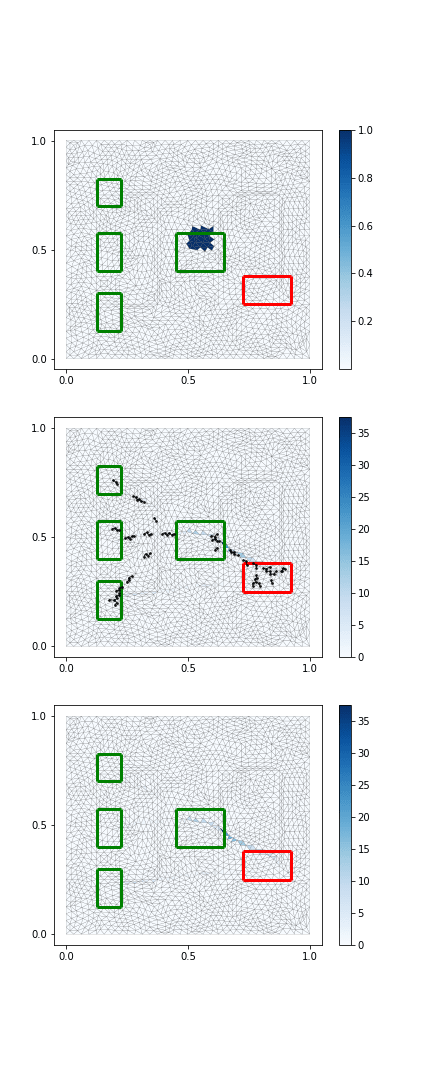}
                \vspace{-2cm}
        \caption{Example 2: delta-like transport density, $\beta = 1.3$}
      \end{subfigure}
        \begin{subfigure}[a]{0.33\textwidth}\label{apxFig:step3}
        \includegraphics[width=\textwidth]{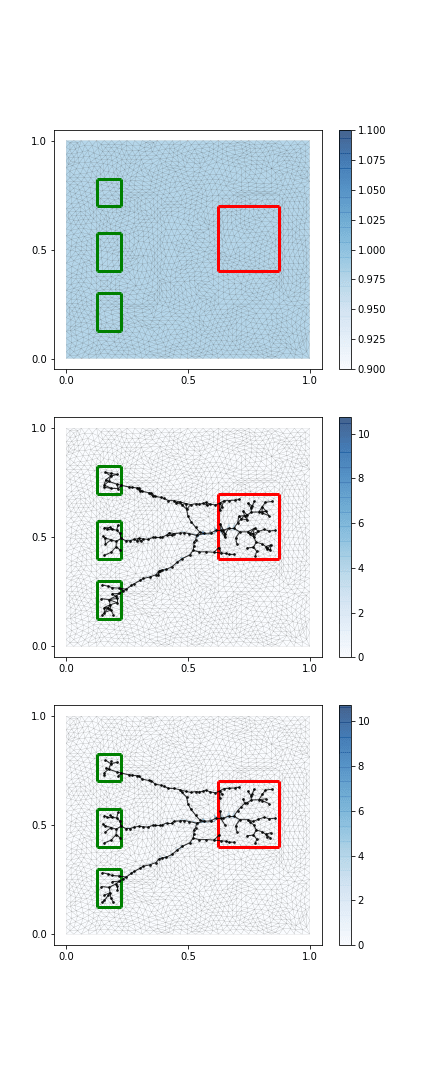}
                \vspace{-2cm}
        \caption{Example 3: uniform transport density, $\beta = 1.5$}
      \end{subfigure}
      \caption{Network extraction examples for branched transportation optimization. Columns denote a particular setup based on source/sink locations and values of $\beta$ and $\mu_{0}$ as described in the sub-captions. Each row represents a step of our protocol: top) initial transport density $\mu_{0}(x,y)$ given in input to \dmkS; center) graph \textit{pre-extraction} using III-AVG (left), II-AVG (center) and I-AVG (right); bottom) graph  \textit{filtering} with weight assigned with rule IBP (left and right) and no simplification (center). The color scheme refers to the value of $\mu$. Sources and sinks are points inside the green and red rectangles, respectively. Note that pre-extraction rule II tends to break the network into disconnected components. This is also the reason why this rule scores well in terms of total path length (see Fig. 4 in the main manuscript). Since the path is already broken and sources cannot reach the sinks, it does not make sense to apply the filtering, hence we left the center-bottom plot empty. }
    \label{apxFig:examples}
\end{figure}

\newpage
\section{Source and sink selection}
\label{apx:source-sink}

\begin{figure}[hptb]
    \centering
    \begin{subfigure}[b]{.85\textwidth}
        \includegraphics[width=\textwidth]{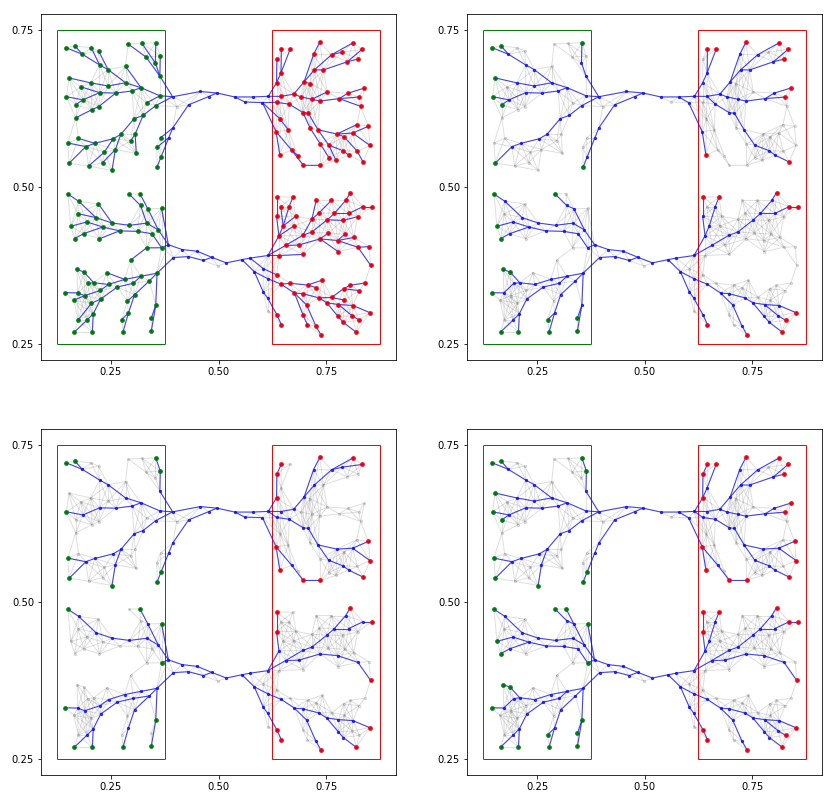}
    \end{subfigure}
      \caption{Choosing sources and sinks. Top-left) no selection of sources and sink ($\tau_{BC}=1$); top-right) selection using betweenness centrality only ($\tau_{BC}=0.01$) ; bottom-left) selection using convex hull-only ($\tau_{BC}=-1$); bottom-right) selection with convex-hull and betweenness centrality criteria ($\tau_{BC}=0.01$).}
      \label{apxFig:source-sink}
\end{figure}

Selecting source and sink is an important task during the graph filtering step as explained in the main manuscript Sec. 3.2. Here we give examples showing why this is the case. Specifically, for a particular problem setup, we consider four cases: i) no selection of sources and sinks. This means that we simply consider sources and sinks \textit{all} the points inside the support of the forcing function $f(x)$, input of the \dmkS. In this case, the final network structure is dominated by the many branches connecting the sources and sinks within the support of $f(x)$, thus hindering the contribution of the bulk of the network, the one connecting sources and sinks; ii) convex hull-only: we set as source (or sink) all the points inside the convex hull of the set of eligible sources (or sinks); iii) betweenness centrality-only:  we set as source (or sink) all the eligible sources (or sinks) that have betweenness centrality smaller than a threshold $0\leq \tau_{BC}\leq 1$. These two criteria miss possibly relevant sub-branches, as shown in Fig. \ref{apxFig:source-sink} top-right and bottom-left; iv) selection using both convex-hull and betweenness centrality criteria. In this case, we capture the relevant sub-branches from both criteria, as shown in the bottom-right of Fig. \ref{apxFig:source-sink}. In the main manuscript we always consider this final criterion as it seems to capture all the relevant branches inside the support of $f(x)$. However, we encourage the final user to choose what criteria to choose based on the application.  Figure \ref{apxFig:source-sink} should help driving this decision. 

\section{Network extraction from image}
\label{apx:net-image}

We describe here the steps followed to get the network shown in Fig. 7 in the main manuscript\footnote{We used image ``IMG\textunderscore 0379.jpg''  stored at KIST\textunderscore Europe\textunderscore data \textunderscore set/raw\textunderscore images/ motion16/ inside the SMGR repository \cite{dirnberger2017introducing}. The dimension of the original one is 5184x3456 pixels. We used a 1200x1200-pixel section in Fig. 7. The coordinates of the bottom-left vertex of the box are $(2000,2000)$ (assuming $(1,1)$ to be the bottom-left pixel of the original image).}.  The original figure contains loops, however our protocol in its standard settings with filtering can only generate tree-like structures. Therefore, if we want to obtain a network with loops, we should consider a minor modification of our routine, as explained in detail below. In short, we need to find first a tree-like network close to the original image with loops and then give this one in input to our routine. Then, after applying our protocol, we can obtain loops by adding edges that properly \textit{close} the loops in the tree. This can all be done in an automatized way as explained below. Otherwise, if obtaining loops is not required, our routine can be used with no modifications.   The whole procedure can be divided into 6 parts: image selection and approximation, graph pre-extraction, tree reduction, terminal identification, graph filtering and edge correction. 

\begin{enumerate}
    \item Image selection and approximation: this consists in choosing an image and, if desired, mapping it into a reduced version of it. This reduction is useful when dealing with high-resolution images.  We use the \textit{resize} function included in \textit{OpenCV}. The pixel values for the reduction image can be computed using different interpolation methods. We use linear interpolation (\textit{cv.INTER\_NEAREST}). The original image's dimensions are 1200x1200. The reduced ones are 100x100. 
    \item Graph pre-extraction: the RGB values of the pixels are mapped into an integer in a one-to-one way, and then they are used to define the function $\mu^*$ on each pixel. The graph $G$ is obtained using rule I-ER as defined in Sec. 2 of the main manuscript.
        \item Tree reduction: the filtering could be applied directly on $G$ after choosing some sources and sinks, but this will generate a filtered tree-like network, i.e., no loops from the image will be captured. Thus, we propose to change $G$ by a tree graph whose structure could then be easily "corrected" (after the filtering step) to get the mentioned loops. This tree is taken to be the \textit{Breadth First Search} graph ($G_{BFS}$) of $G$ from a random root. 
    \item Terminal identification: terminals (the union of sources and sinks) are defined using \textit{filters}, i.e., squared pieces of the image. We place terminals in the graph by counting how many intersections are between the image and the boundary of the filter: if just one intersection is found, then a terminal is placed in the node with the lowest closeness centrality; if three or more intersections are found, then a terminal is placed in the node with the highest closeness centrality. We cover the whole image by disjoint filters, thus we do not miss the relevant parts of the image. Notice that if some parts of the image do not have a representative in one of the two source or sink sets, then they will not be part of the filtered graph. 
    
    \item Graph filtering: $G_{BFS}$ is filtered as in Sec. 3 of the main manuscript (to obtain a graph $G_{f}$) by defining the source set ($S^+$) to be a random element on the set of terminals and the sink set ($S^-$) to be the remaining ones; $\beta_d=1$ in Fig 7. 
    
    \item Edge correction: to \textit{close} the loops we use the following rule. For each pair of terminals $u,v\in G_{f}$, if the shortest path $P^{uv}_{BFS}$ on $G_{BFS}$ is \textit{long} but the shortest path $P^{uv}$ on $G$ is \textit{short,} then 
    add $P^{uv}$ to $G_f$. The notions of \textit{short} and \textit{long} paths are precisely defined by introducing two parameters $L_{BFS}$ and  $L_{G}$, respectively. We say that a path $P$ is \textit{long}, if $l(P)> L_{BFS}$; and it is \textit{short,} if $l(P)< L_{G}$, where $l$ is the number of nodes in the path $P$; the values of $L_{BFS}$ and $L_{G}$ can be tuned based on the system size at hand.
    
\end{enumerate}



\end{document}